\documentclass[aps,pre,twocolumn,showpacs,floatfix,10pt]{revtex4-1}
\usepackage{color}
\usepackage{graphicx}
\usepackage{amssymb,amsfonts,amsmath}
\usepackage{tikz}

\begin{document}

\title{Axelrod's Model with Surface Tension} 
\author{Bruno Pace}
\affiliation{Bioinformatics Group, Institute for Computer Science,
University of Leipzig, H\"artelstra\ss{}e 16-18, D-04107 Leipzig, Germany}
\affiliation{Instituto de F\'{i}sica, Universidade de S\~{a}o Paulo, Caixa Postal 66318, 05314-970, S\~{a}o Paulo, SP, Brazil}
\author{Carmen P. C. Prado}
\affiliation{Instituto de F\'{i}sica, Universidade de S\~{a}o Paulo, Caixa Postal 66318, 05314-970, S\~{a}o Paulo, SP, Brazil}
\email{bruno@bioinf.uni-leipzig.de}
\email{prado@if.usp.br}

\begin{abstract}
In this work we propose a subtle change in Axelrod's model for the dissemination of culture. The mechanism consists of excluding non-interacting 
neighbours from the set of neighbours out of which an agent is drawn for potential cultural interactions. Although the alteration proposed does not
alter topologically the configuration space, it yields significant qualitative changes, specifically the emergence of surface tension, driving 
the system in some cases to metastable states. The transient behaviour is considerably richer, and cultural regions have become stable leading 
to the formation of different spatio-temporal structures. A new metastable ``glassy'' phase emerges between the globalised phase and the polarised, 
multicultural phase.
\end{abstract}

\maketitle

\section{Introduction}

The interest in phenomena involving information flow on networks has been increasing over the last decades. A variety of 
mathematical techniques have been combined to model brain activity, biochemical networks, excitable media, ecosystems and the like \cite{barthelemy}. Naturally amongst 
them are the social systems, including human societies and culture. In an attempt to escape from reductionism, and inspired by the successes 
of statistical mechanics, many different approaches are being proposed to understand the collective behaviour that emerges from the interaction 
of many subparts that constitute the complex system under consideration. There is already a considerable amount of literature on social 
modelling, from applause synchronisation to disease spreading, opinion formation and urban spatial segregation. For a general review, refer to 
\cite{castellano}. In this context, 
Axelrod proposed in 1997 \cite{axelrod} a discrete vector model for cultural dissemination that has been attracting scientists for its very 
simple description but rich behaviour.

Axelrod's model fits in a broad class of models based on agents, where each agent has an internal state and a set of rules determine the 
mechanisms of interaction with other agents. Although very simple in Axelrod's case, this framework is very powerful and its possibilities of 
representing multilevel structures and capturing the essence of quite different systems are vast. In a certain way, Axelrod's model extends some of 
the basic models proposed in the previous decades, such as the voter model, to a discrete vector representation of the internal state of each 
site (or node) of the network. Within this description, agents can be compared by overlapping their vectors. Axelrod's basic premise comes from the 
fact that more similar individuals tend to exchange culturally more often, usually referred to as homophily. Though one agent gets 
more similar to its neighbour after the interaction (local convergence), the global state of the cultural network may in some cases tend to a 
polarised configuration, where groups of different cultures form.   

Probably one of the most important characteristics of Axelrod's model is its out-of-equilibrium phase transition, capturing two different regimes: 
one leading to a globalised state and the other to a polarised one, depending on the model's parameters. Conversely, not much attention has 
been given to its transient behaviour and how the spatial structures develop over time. In fact, although homophily has been the source of 
inspiration for Axelrod's mechanics, during time evolution the information flows through space - cultures hop 
through agents quite freely - and cultural borders in general tend to break, as if people from the same culture did not tend to stick together. To 
account for a stronger homophily we should think of a surface tension effect, for which we propose a mechanism, that is not directly related to a 
memory or majority rule, but on the contrary, emerges naturally from a slight and justifiable change in the updating rules of the model.  

In this work we will outline an analysis of the effect of introducing a sort of agent optimisation, in the sense that agents only use their time 
to interact with whom the interaction is possible. In the next section, we will shortly describe Axelrod's original model and some important 
quantities to characterise the macroscopic state of the system. In the following section we introduce our slight variation in the original 
dynamics and the motivation for doing it. The subsequent section contains a description of the difference attained both in transient and in 
stationary behaviour as well as some results from numeric simulations. Finally in the last section we will discuss the resulting differences and 
present some of the new issues that have arisen from the introduction of surface tension.

\section{Axelrod's Model}

In the original paper \cite{axelrod}, Axelrod defined culture - in a simple fashion - as the set of people's characteristics subject to other 
individuals' influence. In this model, every agent $i$ is represented by a discrete vector 
$\sigma_i = (\sigma_i^1, \sigma_i^2, \ldots, \sigma_i^f, \ldots, \sigma_i^F)$ of $F$ components (or cultural features), each of which can express 
one of $q$ possible cultural traits ($\sigma_i^f \in \{0, 1, \ldots, q-1\}$). Each feature denotes one cultural aspect, such as musical preferences, 
sports, and so forth, and the cultural traits can be thought of as different expressions of these features. The fact that the number of possible 
traits per feature is the same for all features is just a consideration of the model to make it simpler.

The dynamic of the model is driven by the empirical observation that similar individuals are more likely to interact than dissimilar ones 
(homophily). The agents are situated in the nodes of a network. In this work, we have studied the case of a square lattice with periodic boundary 
conditions, but other topologies have been used \cite{transitionscomplex}, \cite{dimensionality}. This is a discrete 
stochastic and asynchronous model. In every time step one agent $i$ is randomly selected along with one of its neighbours $j$. The set of neighbours of
a node $i$ (its neighbourood) is denoted as $\nu_i$. An interaction 
occurs with probability $\omega_{ij}$ equal to their similarity - the similarity between two vectors $\sigma_i$ and $\sigma_j$ is calculated as 
their overlap normalised by the number of features $\omega_{ij} = \frac{1}{F}\sum_{f=1}^F\delta_{\sigma_i^f,\sigma_j^f}$, where 
$\delta_{\sigma_i^f,\sigma_j^f}$ is the Kronecker delta - and it consists of randomly 
selecting one of the features ($f$) whose trait is different between the agents and making the agent $j$ adopt $i$'s trait for that feature after the 
interaction, that is, $\sigma_j^f=\sigma_i^f$. It is worth noting that, in Axelrod's model, it does not matter whether $i$ transfers one trait to $j$ 
or the other way around, it is symmetryc with respect to that. 

Thus, two agents $i$ and $j$ are able to interact if their similarity lies in the range $0<\omega_{ij}<1$, and we will call this link \emph{active}. 
If $\omega_{ij}=0$, their similarity is null and so is their probability of interacting, according to the definition. If $\omega_{ij}=1$, there 
is no more cultural difference between them, and no more possible information exchange. In both cases, the link between these two agents is said 
to be \emph{inactive}. One site (agent) is said to be \emph{active} if it has at least one neighbour with whom the link is active.

\begin{figure}[hbtp]
  \begin{center}
    \includegraphics[scale=0.32]{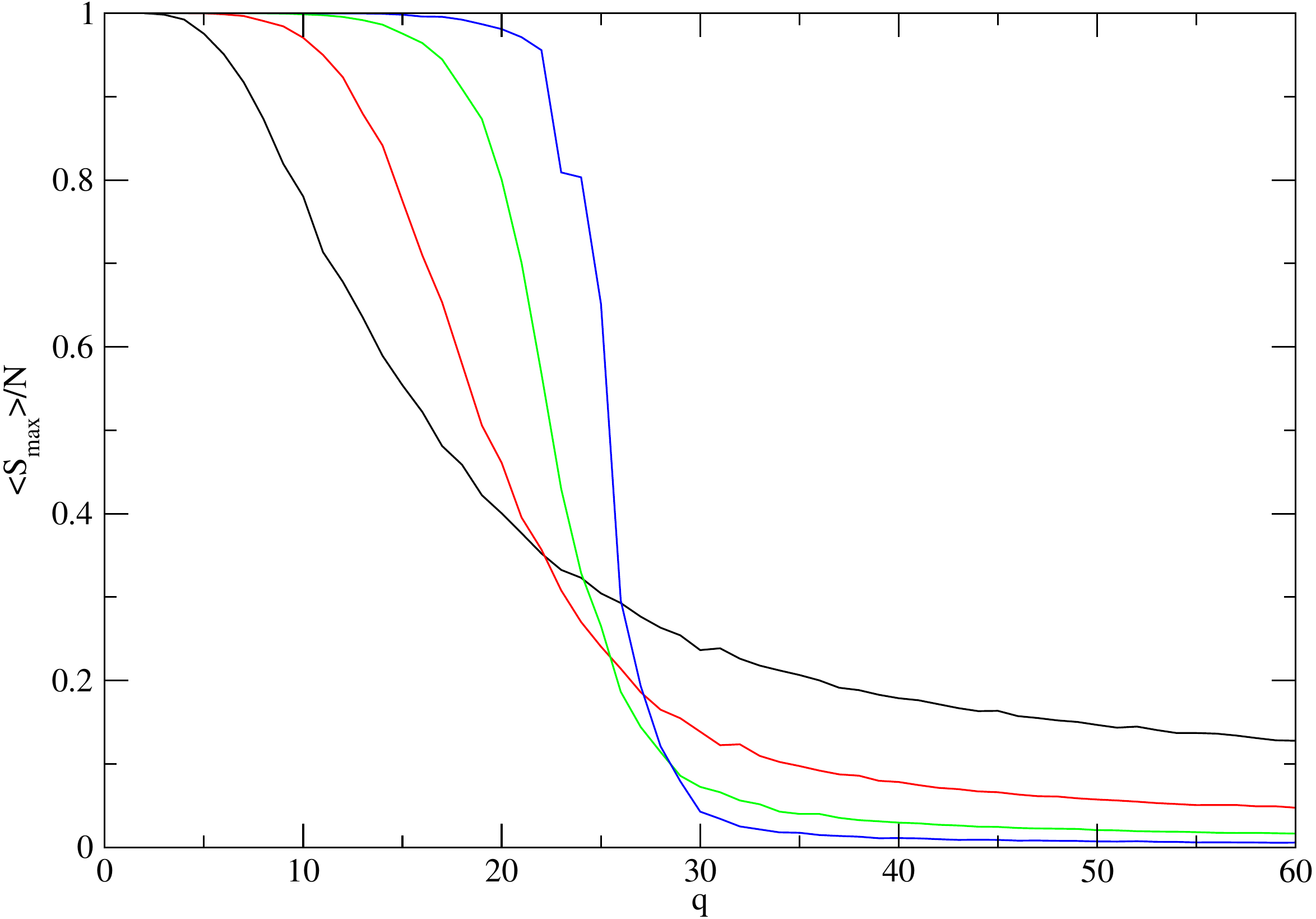}
  \end{center}
\caption{$\frac{\langle S_{max}\rangle}{N}$ vs. $q$ in the absorbing states for several lattice sizes. The curves represent $L=5$ (black), $L=10$ (red), 
$L=20$ (green) and $L=40$ (blue), averaged over $50$, $500$, $1000$ and $2000$ independent realisations, respectively. $F$ was set to $5$.}
\label{Smax}
\end{figure}

Under these rules the system always orbits to an absorbing state (without any more possible interactions) which can - depending on the two 
parameters $F$ and $q$ - be a globalised or a polarised absorbing state. The globalised (monocultural, ordered) state is characterised by the majority of 
agents sharing the same culture (a giant component of one culture), whereas the polarised (multicultural, disordered) state is characterised by many local 
cultural clusters but no global dominant culture. On most topologies, a first order (except for the special case $F=2$) out-of-equilibrium phase transition 
separates these regimes: there is a critical line in the parameter space $q_c(F)$ separating these two different phases. Several order parameters have been 
proposed to characterise the transition, but the most widely used is the average fraction of the dominant culture's size, 
$\frac{\langle S_{max}\rangle}{N}$. As the lattice size increases, 
the transition gets sharper across the critical value (figure \ref{Smax}), 
typical of a first-order transition. It is worth mentioning that the 
(out-of-equilibrium) transition occurs going from one system to another: 
$q$ is part of the definition of the system, and not a collective emergent property such as the temperature. 
The transition happens throughout the 
absorbing states of these different systems as the parameters vary. 

\section{The Model with Surface Tension}

The introduction of surface tension in models on networks is sometimes achieved by the use of a majority rule \cite{kuperman} or a local 
memory \cite{tensionvoter}. In the former case, the configuration space topology is altered - some transitions otherwise possible are no 
longer possible with the majority rule. In this work we propose a subtle modification of the model's dynamics that leads to the emergence 
of surface tension without changing the topology of the configuration space. All the possible transitions are preserved, only the 
probabilities associated with these transitions change.

The main idea behind this modification is a sort of optimisation from the agent's point of view. Suppose - in the Axelrod's case - an agent 
$i$ was selected and it has $4$ neighbours, one of which ($j$) has a state such that $\omega_{ij}=0$ or $\omega_{ij}=1$. Even though there 
is no possible interaction, there is a finite probability ($\frac{1}{4}$) that this neighbour is selected and the iteration will have to be 
discarded. People usually do not try to interact culturally with whom there is no possible interaction, and this is the only alteration introduced 
in the original model. We can now formalise the rules for the model. 

The agents are placed on a network with a specific topology (in our case, a 2D regular lattice with periodic boundary conditions). The concept of 
active sites (or agents) and active links according to the definition in the previous section is still used here. The initial 
conditions are randomly selected from a uniform distribution and the dynamics consist of iterating these two steps:

\begin{description}
  \item[Step 1] Randomly choose one site $i$. Then choose one of its neighbours $j$ such that $0<\omega_{ij}<1$, in case there is (otherwise go back to step 1).
  \item[Step 2] With a probability equal to their similarity, there is an interaction. The interaction consists of randomly 
selecting an $f$ such that $\sigma_i^f\neq\sigma_j^f$ and making site $j$ adopt $\sigma_i^f$.
\end{description}

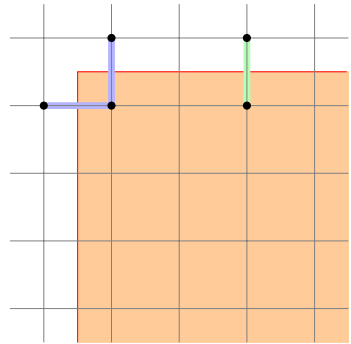
\begin{figure}[ht]
 \centering
 \scalebox{0.3}
{
\begin{tikzpicture}[scale = 3]
  \draw[fill=orange!80!white,opacity=0.5,orange!80!white] (-1.5,-2.5) rectangle (2.5,1.5);
  \draw[very thick][-,shorten >=2pt, color=red]
    (-1.5,-2.5)-- ++(0,4)-- ++(4,0);
  \draw[line width=0.3cm,color=blue!30,cap=round,join=round] (-1,1)--(-2,1) (-1,1)--(-1,2);
  \draw[line width=0.3cm,color=green!30,cap=round,join=round] (1,1)--(1,2);
  \draw[help lines] (-2.5,-2.5) grid (2.5,2.5);
  \fill (-1,1) circle (0.06cm) (-2,1) circle (0.06cm) (-1,2) circle (0.06cm) (1,1) circle (0.06cm) (1, 2) circle (0.06cm);

\end{tikzpicture}
 }
 \caption{Detail of a square border between two cultures, one represented in white and the other in orange. The agents lying on the straight part of the 
interface have the same number of active neighbours on both sides (green), thus being the probability of trait 
transfer the same in both directions. In the corner, however, that symmetry no longer holds (blue). 
There are two agents which can be drawn to transfer their traits to the corner agent, making it twice more probable 
for the corner to be invaded, producing two new corners and continuously ``rounding'' the sharp corners. This 
illustrates the curvature-driven surface tension of the model, not present in Axelrod's original model.}
\label{grid}
\end{figure}

Note that the only difference compared to Axelrod's original model is the constraint that the randomly selected neighbour has $0<\omega_{ij}<1$. This 
choice makes the model asymmetric with respect to the trait transfer direction. If we invert and make site $i$ adopt $\sigma_j^f$ in step $2$, the 
opposite effect arises: the total interface length tends to increase and a physically implausible cultural mixing occurs.  
Suppose now a site $i$ is selected and only one of the four links connecting it to its neighbours is inactive. In this case, there are only three possible 
neighbours to be randomly drawn, with probability $\frac{1}{3}$ for each. After the choice, there is still a probability that no interaction occurs, but 
all the selected pairs may potentially exchange cultural information. This single modification is responsible for the emergence of surface tension 
(see figure \ref{grid}). 

A colour scheme was devised for the simulations in order to observe the spatiotemporal patterns. Different colours were assigned to 
different cultural states. Being the RGB colour space three-dimensional, it is impossible to assign similar colour to similar cultures based on distances. 
Thus, the colour to culture mapping does not indicate that similar colours represent similar cultures.

\section{Numerical Results}

\begin{figure}[hbtp]
  \begin{center}$
    \begin{array}{cccc}
    \includegraphics[keepaspectratio=true, width=2cm]{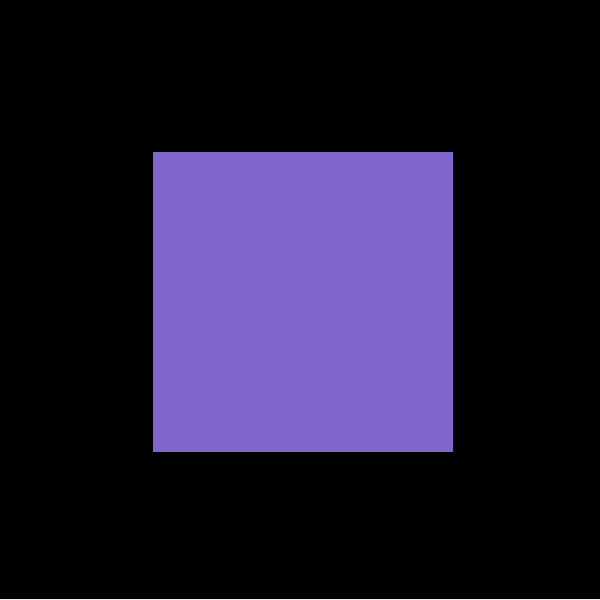}& 
    \includegraphics[keepaspectratio=true, width=2cm]{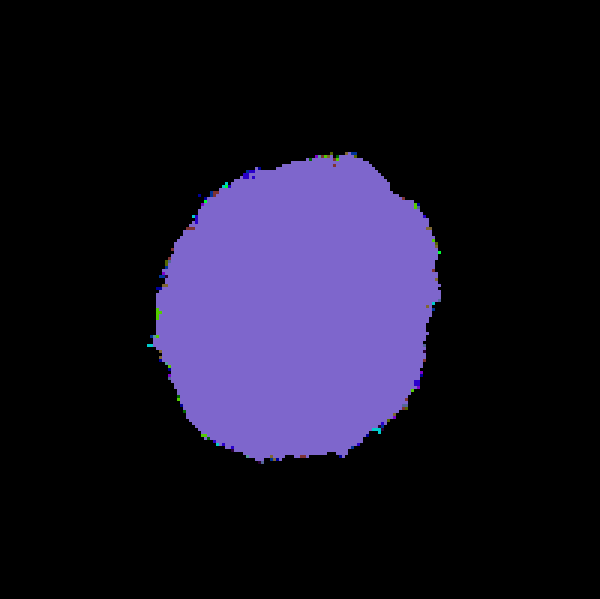}&
    \includegraphics[keepaspectratio=true, width=2cm]{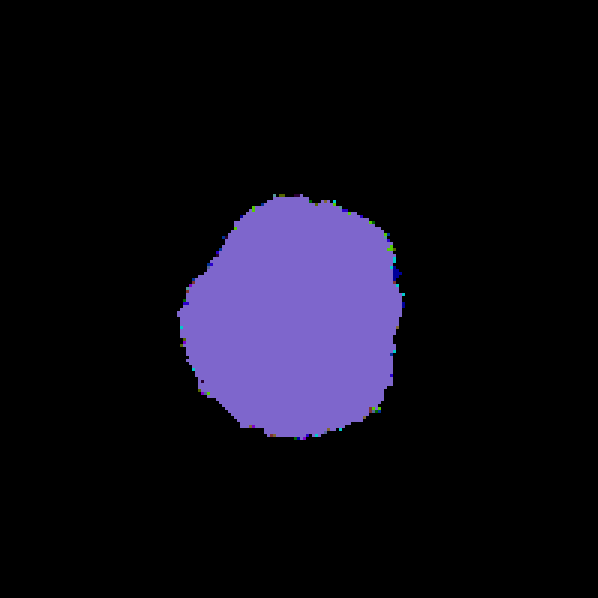}& 
    \includegraphics[keepaspectratio=true, width=2cm]{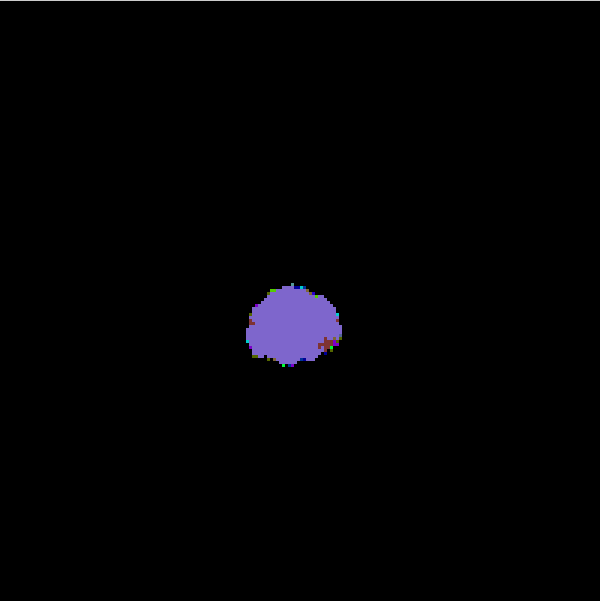}\\
    \includegraphics[keepaspectratio=true, width=2cm]{quadrado.png}& 
    \includegraphics[keepaspectratio=true, width=2cm]{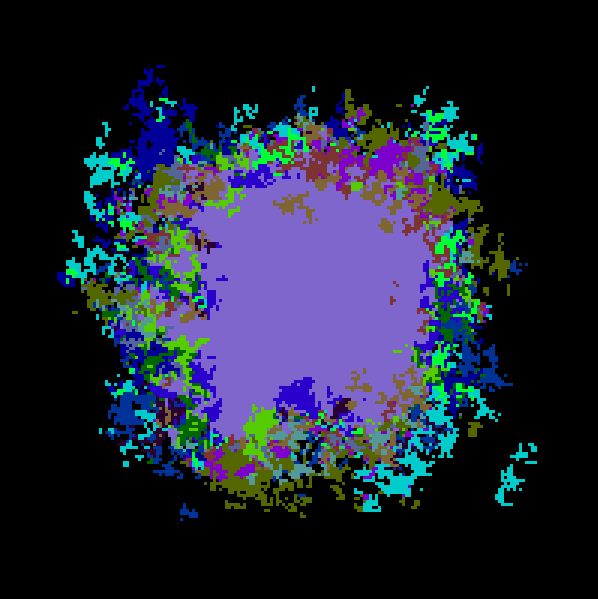}&
    \includegraphics[keepaspectratio=true, width=2cm]{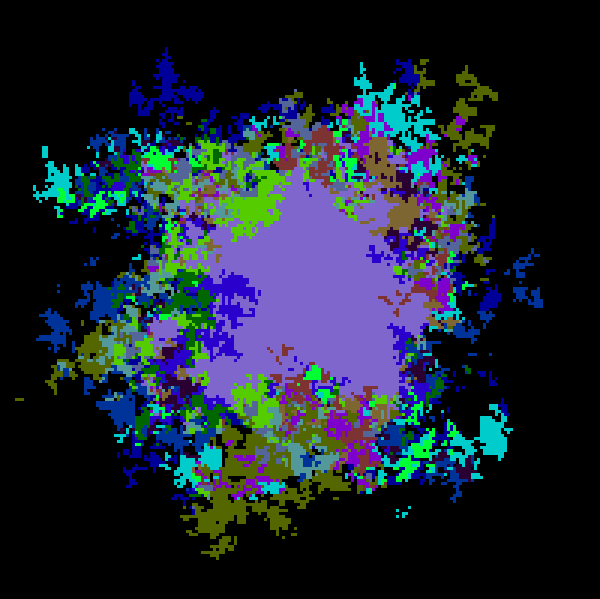}& 
    \includegraphics[keepaspectratio=true, width=2cm]{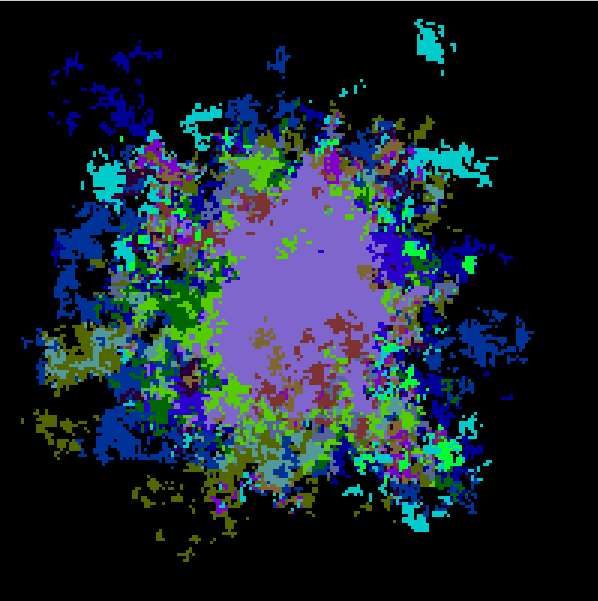}
  \end{array}$\\
 \end{center}
\caption{Same initial conditions (square), the first row corresponding to the case with surface tension and the second to the Axelrod's original model.}
\label{bubble}
\end{figure}

\begin{figure}[hbtp]
  \begin{center}$
   \begin{array}{cc}
    \includegraphics[keepaspectratio=true,width=3cm]{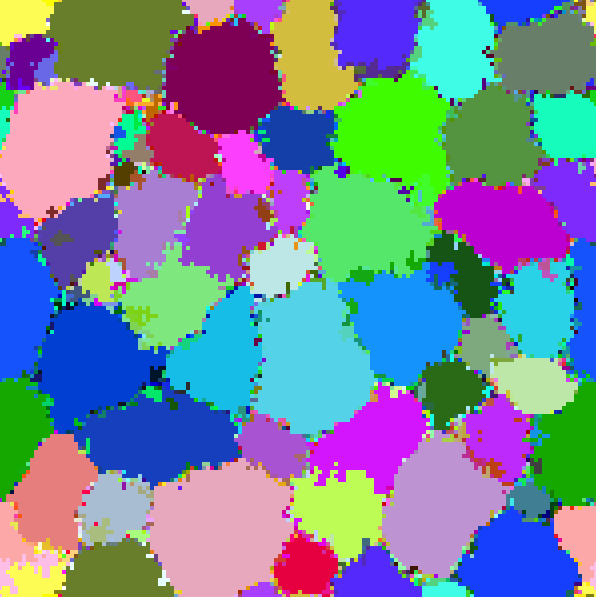}&
    \includegraphics[keepaspectratio=true,width=3cm]{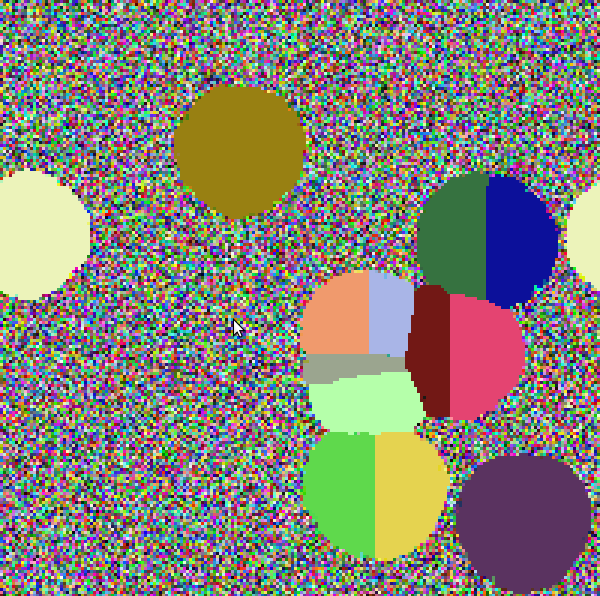}
  \end{array}$
 \end{center}
\caption{Configuration during transient for (left) $F=11$ and $q=2$, showing the homogeneous nucleation of bubbles with non-rigid membranes and 
multiculturality on the borders between different regions and (right) $F=25$ and $q=10$, showing the non-homogeneous nucleation of bubbles, which will 
grow and reach a metastable state.}
\label{multicultural}
\end{figure}

While seemingly subtle, there are some radical changes in the overall properties of the model. In Axelrod's model, the absence of surface tension 
leads to instability of cultural regions, and bubbles tend to break \cite{compsimul} (figure \ref{bubble}). In our model, cultures have a tendency to organise 
in space, forming well-defined cultural regions throughout the trajectory in time. This can be interpreted as cultural cohesion, or a tendency of 
people from the same culture to stick together: the principle of homophily in a stronger sense. As a consequence, instead of local cultural 
interfaces there are spatially extended interfaces between cultures, macroscopic membranes, which usually do not break. These membranes have very 
peculiar interacting mechanics, and a more careful investigation is yet to be made. The isolation of these membranes is possible exporting these rules to 
the voter model, according to \cite{ourvoter}. Another remarkable consequence of the formation of domains is that the cultural exchange happens in the 
borders between regions, and the borders tend to concentrate cultural diversity (see figure \ref{multicultural}). 

The transient regime with Axelrod's original dynamics has in general no tendency for cultural spatial 
structures to form in time. In contrast, after introducing surface tension it is possible to observe some different spatiotemporal patterns. For small 
enough values of $q$ - the region of the parameter space where the globalised absorbing state is reached - the cultural exchange rapidly generates a 
homogeneous nucleation, covering the whole lattice with local polygonal cells whose borders (or membranes) are flexible enough, allowing bigger ones to 
coalesce progressively, becoming even bigger at the expense of anihilating the small ones (figure \ref{multicultural}). For large enough values of $q$, there are some nuclei 
with possible culture exchange immerse in a vast disordered ocean of cultural disagreement. In this region, the transient is very similar to 
Axelrod's, despite the formation of membranes within the islands of cultural exchange. The absorbing state reached is also similar to 
Axelrod's in this case. 

Useful to describe the macroscopic state are the densities of links, divided in three categories: (inactive) links 
between nodes with no feature in common, (active) links between nodes with some features - but not all - in common, and (inactive) links between 
nodes with the same cultural state (all features in common). The densities of these links will be called $\rho_0$, $\rho_a$ and $\rho_F$ 
respectively, and are given by the following expressions on a regular lattice with periodic boundary conditions and $N$ nodes.

\begin{equation*}
\rho_0 = \frac{1}{4N}\sum_{i=1}^N\sum_{j\in\nu_i}\delta_{\omega_{ij},0}
\end{equation*}

\begin{equation*}
\rho_F = \frac{1}{4N}\sum_{i=1}^N\sum_{j\in\nu_i}\delta_{\omega_{ij},1} 
\end{equation*}

\begin{equation*}
\rho_a = 1 - \rho_0 - \rho_F
\end{equation*}

\begin{figure}[hbtp]
  \begin{center}
 \includegraphics[keepaspectratio=true,width=8cm]{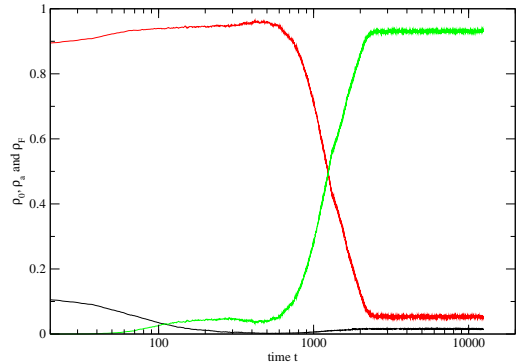}
 \end{center}
\caption{Graph illustrating the transient behaviour or the three categories of bonds: between completely dissimilar sites $\rho_0$ (black), between interacting 
agents $\rho_a$ (red) and between culturally identical agents $\rho_F$ (green), for a lattice size $L = 100$, with $N=10000$ nodes. In this case, $F=25$ and $q=12$.
The self-averaging makes the curve smoother as lattice size grows. A metastable state is reached for this choice of parameters (observe that the density 
of active interfaces does not decrease to zero).}
\label{rho0AF}
\end{figure}

One can see how these quantities develop during the transient until it reaches a metastable state in the figure \ref{rho0AF}.

A function we used to detect metastable configurations is the number of surviving traits, proposed in \cite{remainingtraits}, which is a Lyapunov function 
for both models. It is calculated by counting the traits that are still circulating on the network, and has maximum value $qF$ and minimum value 
$F$. Observe that if the numeric value of two different cultural features is the same, it is counted twice. The fact that they have the same symbol
does not mean anything since they stand for different cultural features, thus being incomparable.

\begin{equation*}
ST = qF - \sum_{f=1}^F\sum_{t=0}^{q-1}[\delta_{0,\sum_{i=1}^N\delta_{t, \sigma_i^f}}]  
\end{equation*}

This function, besides being monotonically decreasing in time, may provide insights into the mechanics of the models. At each interaction, one 
agent adopts one of the cultural traits of its influencing neighbour and discards the previous value for that feature. In this way, there are 
eventual trait extinctions during the system's trajectory in time. Once there is an extinction of a certain trait, it will never get back to the 
network and many configurations previously accessible become inaccessible. Kuperman has also studied two different Lyapunov functions for his 
model with majority rule \cite{kuperman}, and the use of these functions to understand stability \cite{stability} or detect metastable 
configurations are of great importance.

What is entirely new is a metastable ``glassy'' phase for intermediate values of $q$. The transient regime is characterised 
by the non-homogeneous nucleation and growth of bubbles (figure \ref{multicultural}), and the nucleation usually occurs at neighbour sites with orthogonal 
cultures (in the case of non-periodic boundary conditions, the edges and corners become nucleation regions). Therefore, the average density of nucleation 
regions also varies with $q$. After growing, these bubbles eventually touch each other, forming well defined domains whose borders are almost 
completely rigid; yet, there might still be some occasional changes. In case of dense nucleation, the mean domain size reached is smaller and the domains 
become more susceptible to trait invasion from neighbour ones, provoking some extra domain concatenation and leading to ultimate irregular domain shapes. 

\begin{figure}[hbtp]
  \begin{center}$
    \begin{array}{ccc}
    \includegraphics[keepaspectratio=true, width=2.6cm]{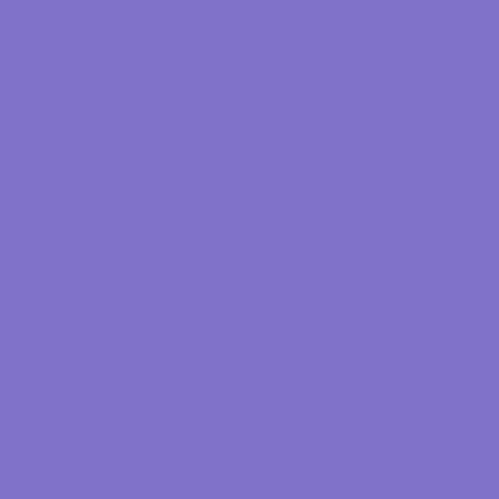}&
    \includegraphics[keepaspectratio=true, width=2.6cm]{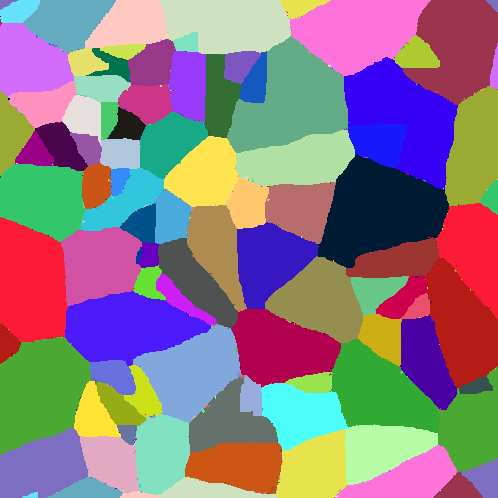}& 
    \includegraphics[keepaspectratio=true, width=2.6cm]{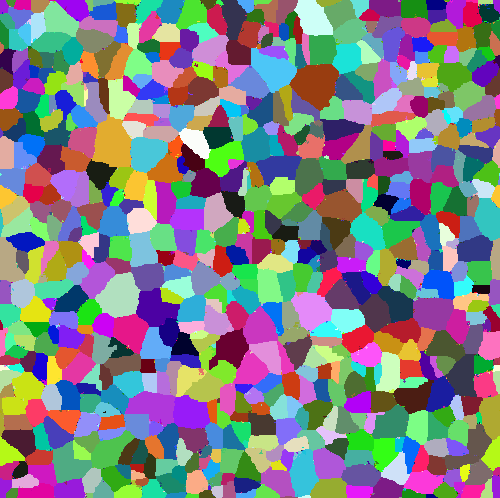}\\
    \includegraphics[keepaspectratio=true, width=2.6cm]{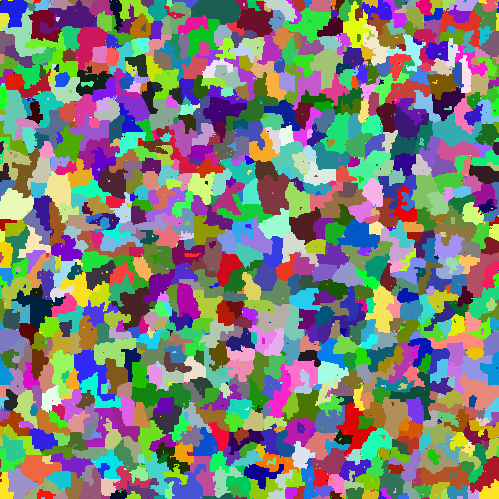}& 
    \includegraphics[keepaspectratio=true, width=2.6cm]{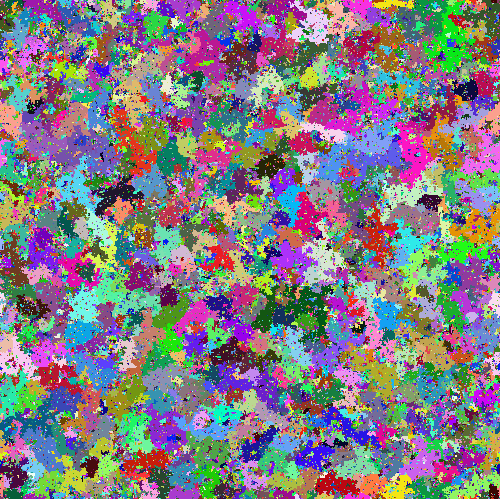}&
    \includegraphics[keepaspectratio=true, width=2.6cm]{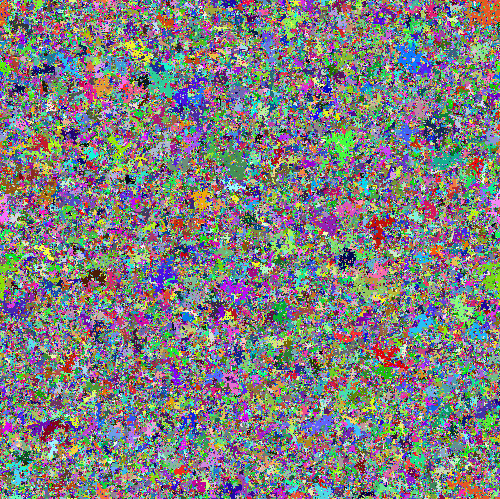}
  \end{array}$\\
 \end{center}
\caption{Absorbing states or metastable states for choices of parameters lying in the three different phases. $F=25$ and $L=500$ for all of them.
First row: $q=2$, $q=10$ and $q=20$. Second row: $q=50$, $q=100$ and $q=125$.}
\label{transition}
\end{figure}

These states are metastable in the sense that they are not absorbing states (there 
are still cultural exchanges taking place), but are quasi-stationary, remaining confined in certain attractor-like regions of the configuration space. 
In \cite{stability} a completely different notion of metastability is used, not to be mistaken with the one used here.
It is important to note that what defines these regions as attractors are the transition probabilities, and not the topology: if we 
use these metastable configurations as the initial conditions for the Axelrod's case, these domains break, and the consensus state is reached. 
Under Axelrod's dynamics, there are only fixed point attractors, but in our case there are other kinds, characterising a different attractor structure, 
albeit the common topology. The absorbing states or metastable states characterising the three different phases are represented in figure \ref{transition}.

The next step would be to characterise both transitions and find the critical lines in the parameter space. This however is not such a simple 
task. Using the typical order parameter $\frac{\langle S_{max}\rangle}{N}$ apparently gives us a good initial idea about the critical values but 
when the lattice size increases, the middle plateau (corresponding to the metastable phase) progressively lowers, tending eventually to zero. 
We have also tried to use the cluster entropy proposed in \cite{temperature}, which works pretty well for Axelrod's original model 
but not for ours with surface tension. One of the difficulties of dealing with this out-of-equilibrium transition involving metastable states is 
deciding when to interrupt the simulation to measure the order parameter of that (truncated) configuration, or even to understand the difference 
between long transients or metastable states.


\section{Discussion}

Axelrod's model is a simplification of a very complex problem, and it was not proposed in an attempt to describe or explain the intricate process 
of cultural exchange and formation. That said, the model is interesting in itself, and that probably is the great motivation for studying it. 
According to our view, one of its main limitations is related to the absence of surface tension during its time evolution. It is possible to observe 
that cultural traits flow through the agents in a way that the cultural borders tend to break, forming very different spatiotemporal patterns from 
what we expect due to the homophilic nature of social interactions - affine agents staying together in space. In Axelrod's model, cultural 
bubbles are however unstable and usually dissolve or break, depending on the surroundings. Besides that, both models still lack a mechanism for creation 
of novelty, since the traits only get extinct in the course of time. 

We proposed a modification in the rules that results in surface tension without changing topologically the configuration space, 
clearly implying a modification of the transient behaviour, making it much richer: cultural borders do not break anymore, cultural regions 
become connected in space, spatial patterns emerge such as bubble nucleation and surface-tension-driven polygonal cells. The new model 
still preserves the globalised and polarised phase of Axelrod's model, but a new ``glassy'' metastable phase emerges between them, 
with cultural regions resembling political maps whose (active) borders concentrate all the cultural exchange and diversity. Unlike Axelrod's model, this 
phase has the noteworthy characteristic of not reaching a frozen state. This novel outcome accounts for social interactions in a 
more realistic way, with spontaneous formation of cultural niches and, more importantly, the boundaries that separate and define them.

An important issue to be pointed out is the dependence of system's evolution on the uniform random distribution as the initial 
condition. The polarised phase actually stems from this random initial condition in situations where $q$ is large enough such that it is quite 
improbable that any two neighbours have something in common. Much of our understanding of the model comes from this widespread choice, but some 
carefully chosen initial conditions lead to significantly different situations that must be understood. We have used some of these particular 
configurations as initial conditions to unravel the mechanics of the membranes, and they can be quite useful.

\section{Acknowledgements}

We thank FAPESP and CNPq for their financial support, making this work possible. Many special thanks to Konstantin Klemm, for the useful discussions 
during the preparation of this manuscript.

\bibliography{bibliografia}

\end{document}